\documentclass[lettersize,journal]{IEEEtran}
\usepackage{amsmath,amsfonts}
\usepackage{algorithmic}
\usepackage{array}
\usepackage{textcomp}
\usepackage{stfloats}
\usepackage{url}
\usepackage{verbatim}
\usepackage{graphicx}
\def\BibTeX{{\rm B\kern-.05em{\sc i\kern-.025em b}\kern-.08em
    T\kern-.1667em\lower.7ex\hbox{E}\kern-.125emX}}
\usepackage{balance}
\usepackage{subcaption}
\usepackage{multicol}
\usepackage{multirow}
\usepackage{blindtext}
\begin{document}
\title{Artificial Synapse based on ULTRA\textbf{RAM} Memory Device for Neuromorphic Applications}
%\title{Artificial Synapse based on ULTRA\textbf{RAM} Memory Device for Efficient Deep Neural Networks }
\author{Abhishek Kumar, Peter D. Hodgson, Manus Hayne, and Avirup Dasgupta
	\thanks{Abhishek Kumar is with the Department of Electrical Engineering and Computer Sciences, University of California at Berkeley, USA (Email: abhishekg@berkeley.edu). ORCID: 0000-0002-9355-3354. }
	\thanks{Avirup Dasgupta is with the Department of Electronics and Communication Engineering, Indian Institute of Technology Roorkee, Roorkee 247667, India (Email: avirup@ece.iitr.ac.in). }
	%\thanks{Peter D. Hodgson and Manus Hayne are with the Department of Physics, Lancaster University, Lancaster LA1 4YB, United Kingdom.}
	\thanks{Peter D. Hodgson and Manus Hayne are with Quinas Technology, Lancaster LA1 4YB, United Kingdom; and the Department of Physics, Lancaster University, Lancaster LA1 4YB, United Kingdom (Email: p.hodgson1@lancaster.ac.uk, m.hayne@lancaster.ac.uk).}
	\thanks{Corresponding author(s): Abhishek Kumar and Avirup Dasgupta}
%	\vspace{-6ex}
}

\markboth{Authors' pre-publication version 2025. This version has not been peer reviewed.}%
{}

\maketitle

\begin{abstract}
	The memory demands of large-scale deep neural networks (DNNs) require synaptic weight values to be stored and updated in off-chip memory like dynamic random-access memory, which reduces energy efficiency and increases training time. Monolithic crossbar or pseudo-crossbar arrays using analog non-volatile memories, which can store and update weights on-chip, present an opportunity to efficiently accelerate DNN training.
	In this article, we present on-chip training and inference of a neural network using an ULTRA\textbf{RAM} memory device-based synaptic array and complementary metal-oxide-semiconductor (CMOS) peripheral circuits.
	ULTRA\textbf{RAM} is a promising emerging memory exhibiting high endurance ($>$$10^7$ P/E cycles), ultra-high retention ($>$$1000$ years), and ultra-low switching energy per unit area. 
	A physics-based compact model of ULTRA\textbf{RAM} memory device has been proposed to capture the real-time trapping/de-trapping of charges in the floating gate (FG) and utilized for the synapse simulations. A circuit-level macro-model is employed to evaluate and benchmark the on-chip learning performance in terms of area, latency, energy, and accuracy of an ULTRA\textbf{RAM} synaptic core.
	In comparison to CMOS-based design, it demonstrates an overall improvement in area and energy by 1.8$\times$ and 1.52$\times$, respectively, with 91\% of training accuracy. 
\end{abstract}
\begin{IEEEkeywords}
	ULTRA\textbf{RAM}, Non-volatile Memory, Compound Semiconductor, DRAM, Flash
\end{IEEEkeywords}
%
%============= New section ==================
\section{Introduction}
\label{sec:introduction}
%
%============== TEM Images ==========================
%\begin{figure}
%	\centerline{\includegraphics[width=1\columnwidth]{./images/TEM-Si-substrate.jpg}}
%	\caption{ III-V on Si material characterization. (a) Dark-field g = 220 transmission electron microscope (TEM) image of a GaSb/Si buffer layer. (b) Dark-field g = 002 TEM image of the ULTRA\textbf{RAM} sample \cite{n1}. Only the memory layers and the top of the GaSb buffer are visible in this image. A single 60° misfit dislocation is visible in the InAs/GaSb buffer interface.}
%	\label{tem_img}
%\end{figure}
%
%
Deep neural networks (DNNs) have demonstrated remarkable success across various applications, including image classification, speech recognition, time-series prediction, and spatiotemporal recognition tasks \cite{ch8dnn1, ch8dnn3}. However, DNNs implemented on conventional von Neumann computing architectures suffer from significant energy consumption and high latency \cite{ch1r2}. 
This is due to the memory demands of the large-scale neural networks often surpassing the capacity of on-chip SRAM caches \cite{ch8intro3}. Additionally, expanding SRAM size is constrained due to the considerable cell area requirement (100-200F$^2$), making scalability a challenge \cite{ch8intro2, ch8intro4}. As a result, high-bandwidth off-chip memory, such as DRAM, is commonly utilized to store network parameters \cite{ch8dram1}. However, this approach reduces energy efficiency and increases latency compared to on-chip solutions due to the constraints of the von-Neumann bottleneck \cite{ch8dram2, ch5app1}.
%This is due to the memory demands of large-scale DNNs require synaptic weight values to be stored and updated in off-chip memory like DRAM, which reduces energy efficiency and increases training time. 
In a fully connected DNN, training can be significantly accelerated by reducing data movement through on-chip storage and conducting weight updates directly at the same node, with all nodes interconnected within an array.
\par
Monolithic crossbar or pseudo-crossbar arrays using analog non-volatile memories, which can store and update weights on-chip, present an opportunity to accelerate DNN training by reducing data movement \cite{ch8neuro1}. Various emerging non-volatile memory technologies, such as resistive random-access memory (RRAM) \cite{ch8rram1, ch8rram3}, phase-change memory (PCM) \cite{ch8pcram1}, and ferroelectric devices \cite{ch5neuro1, ch8fe3}, are promising candidates due to their compact cell size and capability to store multiple intermediate states.
However, PCM experiences a sudden reset transition, whereas oxygen vacancy-based RRAM devices are prone to cycle-to-cycle variability and limited $G_{max}/G_{min}$ ratios, which leads to asymmetric potentiation and depression characteristics \cite{ch8rram4}. Additionally, the slow write speeds, ranging from microseconds to milliseconds, can significantly prolong training duration, potentially extending to several years \cite{ch5neuro1, ch8neurosim1}.
%
%============== Schematic UR ==========================
\begin{figure}
	\centering
	\includegraphics[width=1\columnwidth]{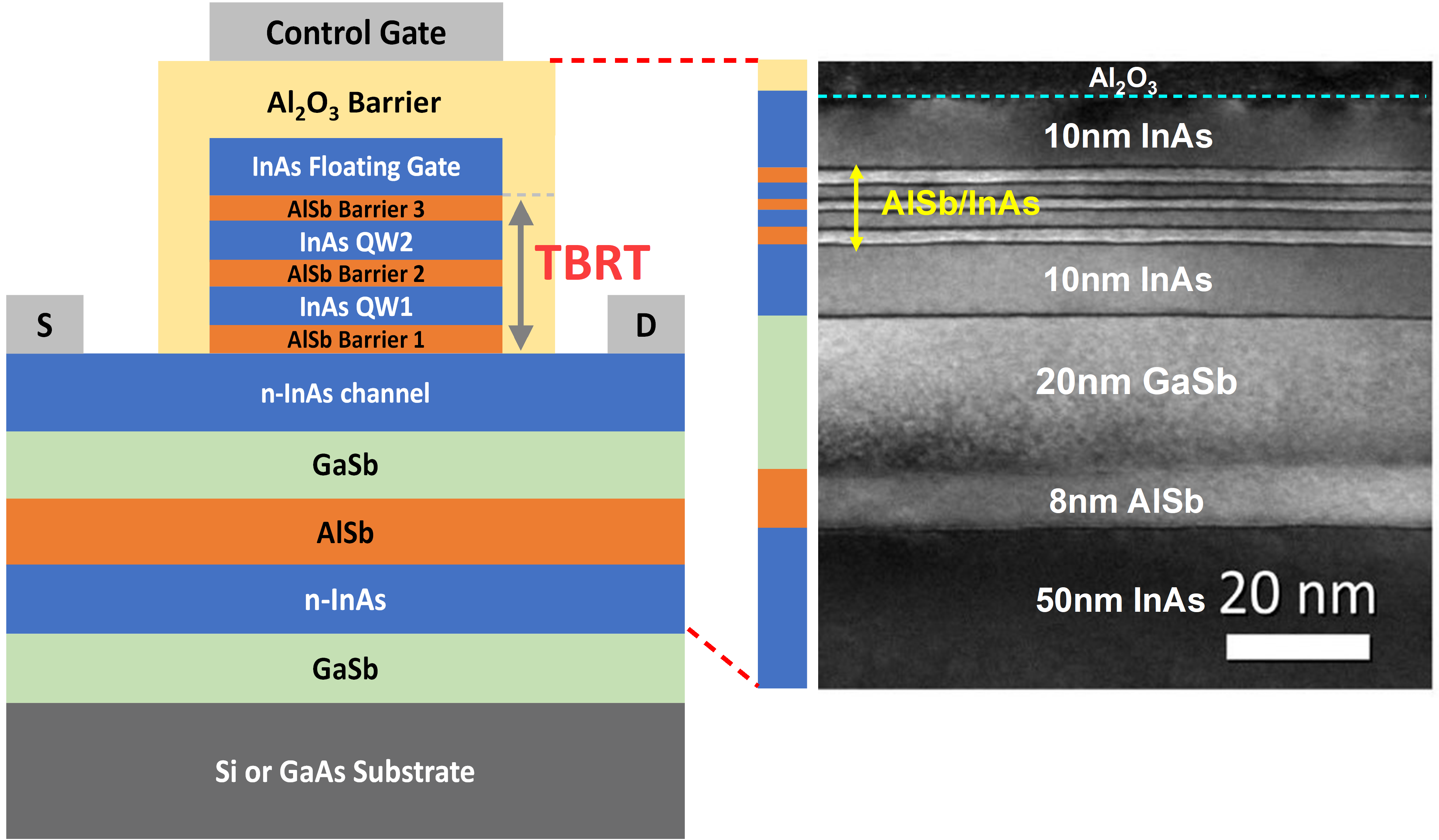}
	\caption{Schematic of an ULTRA\textbf{RAM} memory cell and the corresponding transmission electron microscope image of the device's epilayers \cite{ch7ur1.ted}.}
	\label{schematic_ur}
\end{figure}
\par
In this paper, we present on-chip training and inference of a neural network using an ULTRA\textbf{RAM} memory device-based synaptic array and CMOS peripheral circuits.
A physics-based compact model of an ULTRA\textbf{RAM} memory device has been used to capture the real-time trapping/de-trapping of charges in the floating gate (FG) and utilized for the synapse \cite{mypaper_uredtm, mypaper_urdrc}. A circuit-level macro-model is employed to evaluate and benchmark the on-chip learning performance in terms of area, latency, energy, and accuracy of an ULTRA\textbf{RAM} synaptic core \cite{ch8dnn4}.
In comparison to CMOS-based SRAM design, it demonstrates an overall improvement in area, energy, and latency with 91\% training accuracy. 
\section{Memory Properties of ULTRA\textbf{RAM}}
\begin{figure}
	\centering
	\begin{subfigure}{0.5\linewidth}
		\includegraphics[width=\linewidth, keepaspectratio]{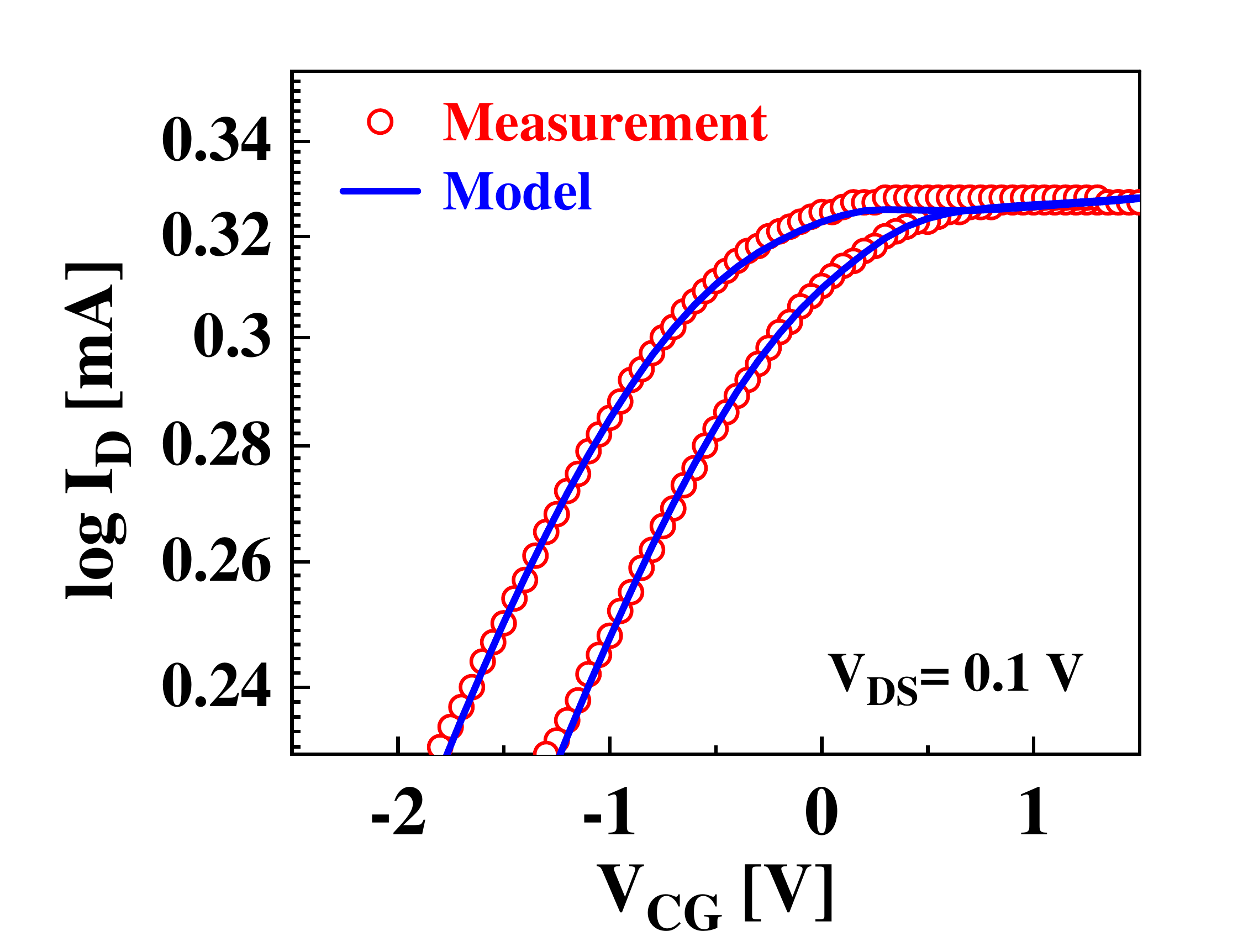}
		\caption{}
		\label{fig_idvg}
	\end{subfigure}\hfil
	\begin{subfigure}{0.5\linewidth}
		\includegraphics[width=\linewidth, keepaspectratio]{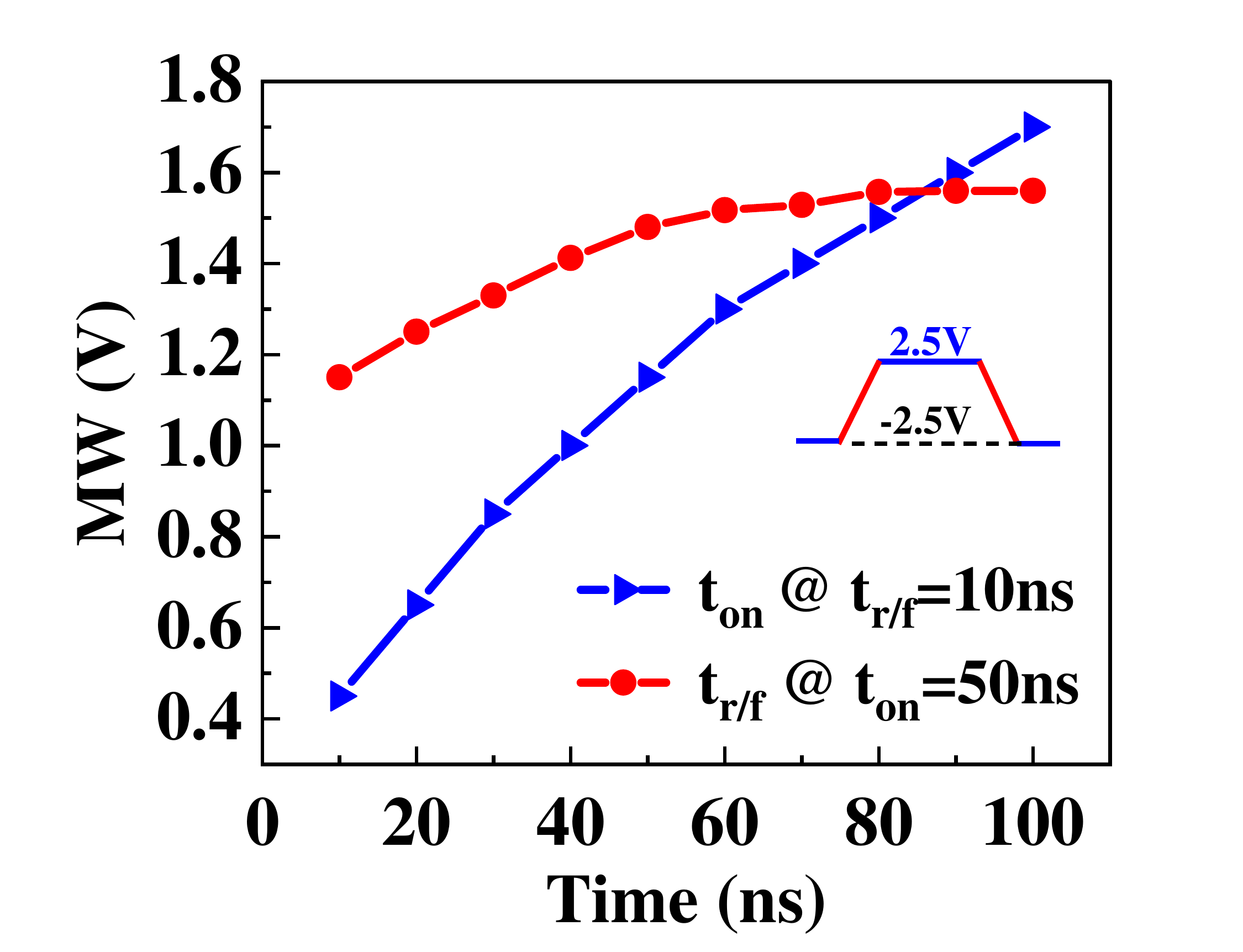}
		\caption{}
		\label{mw_vs_time}
	\end{subfigure}\hfil
	\caption{(a) Validation of model with experimental I-V characteristics \cite{ch7ur2.aem}. (b) Variations in the memory window (MW) of the device for pulse width and rise/fall time.}
	\label{model_fig}
\end{figure}

ULTRA\textbf{RAM} is a promising emerging memory exhibiting high endurance ($>>$$10^7$ P/E cycles\footnote[1]{Experiment limited. Zero degradation observed after $10^7$ program/erase cycles.}), ultra-high retention ($>$$1000$ years), and ultra-low switching energy per unit area \cite{ch7ur2.aem, ch7ur1.ted}. The state is determined by the presence or absence of electrons in a floating gate (FG). Unlike a single SiO$_2$ barrier in flash memory, the novelty comes from the InAs/AlSb triple barrier resonant tunneling (TBRT) structure \cite{ch7tbrt3}, as shown in Fig. \ref{schematic_ur}. TBRT structure provides a high-potential electron barrier with no bias and allows fast resonant tunneling to program/erase pulse ($\pm$2.5V) with switching energy per unit area 1000 times lower than NAND flash, and 100 times lower than DRAM \cite{ch7ur3.edtm}.
A physics-based compact model of an ULTRA\textbf{RAM} memory device has been used to capture the real-time trapping/de-trapping of charges in the floating gate (FG) and utilized for the synapse \cite{mypaper_uredtm, mypaper_urdrc}.
Fig. \ref{fig_idvg} shows the I-V characteristics of an ULTRA\textbf{RAM} cell. The obtained memory window ($MW$ = $V_{th,program} - V_{th,erase}$) depends on the input waveform, which is accurately captured in real time by the proposed model, as shown in Fig. \ref{mw_vs_time}.
%
% ======= EXP IV  ===============
%
\section{DNNs using ULTRA\textbf{RAM} Synapse}
The in-memory computing (IMC) architecture accelerates convolutional-neural-network (CNN) processing by executing matrix-vector multiplications directly within the memory crossbar array. The fundamental concept of analog IMC is to represent weights as conductance states within memory cells, mimicking synaptic behavior. 
In this work, we have utilized an ULTRA\textbf{RAM} memory device as a synapse, which enables the storage of multiple conductance states. First, we have employed experimentally demonstrated ULTRA\textbf{RAM} cells to evaluate the actual on-chip performance. Since the currently fabricated devices have relatively long channel lengths ($\sim$10~$\mu$m) and no other emerging memory technologies are available at this scale, their performance has been compared against conventional SRAM-based synapses to provide a consistent estimation of performance metrics.
Secondly, we have projected the on-chip performance with scaled-down simulated devices that match the current state-of-the-art features sizes of other emerging memory technologies.  
%
% ======= DNN Architecture ===============
\begin{figure}
	\centering
	\includegraphics[width=1\columnwidth]{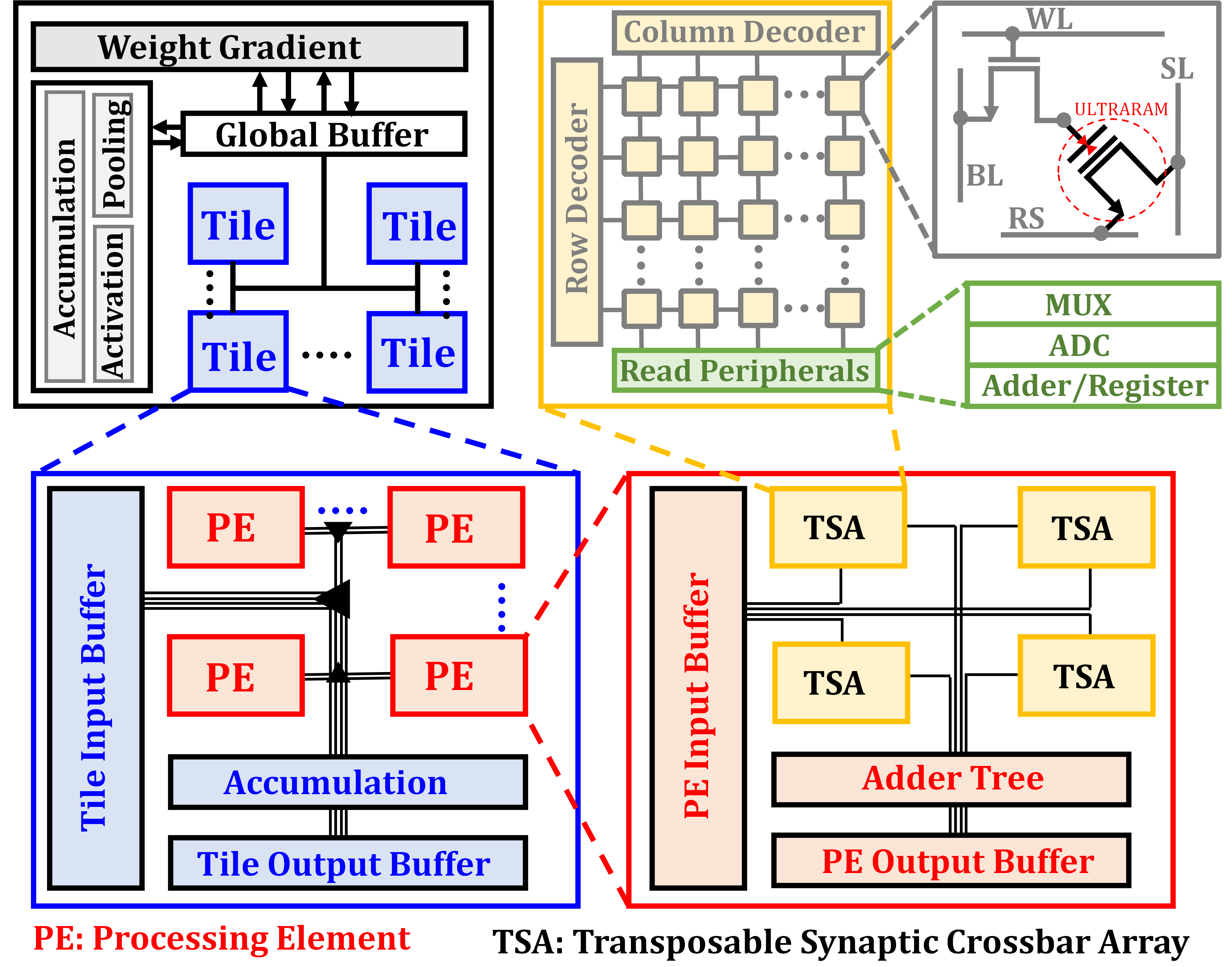}
	\caption{Architecture-level representation of ON-chip learning hardware.}
	\label{dnn_arch}
\end{figure}
%
%
%
% ======= VGG-8 CIFAR-10 dataset ===============
\begin{figure}
	\centering
	\includegraphics[width=1\columnwidth]{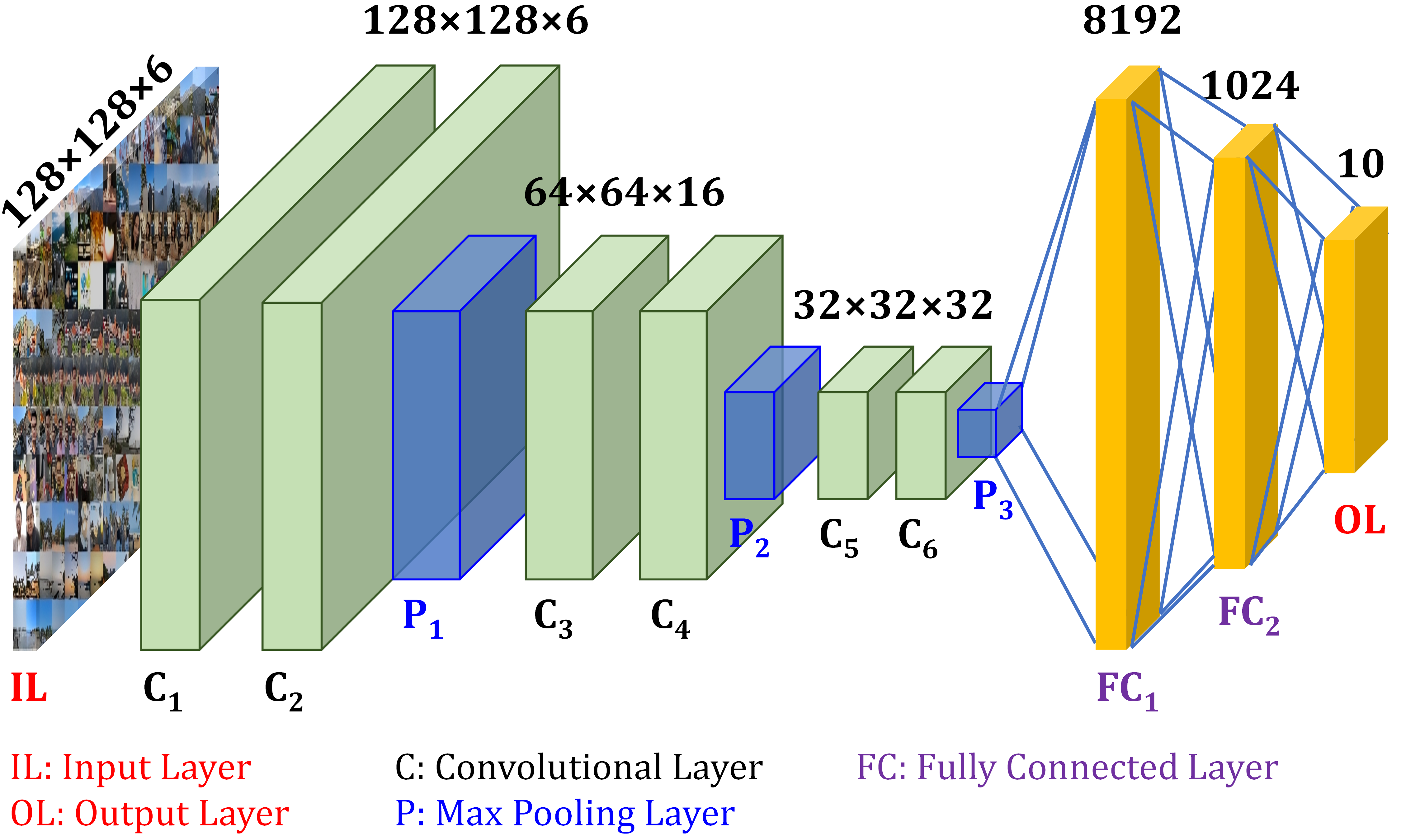}
	\caption{Schematic of the VGG-8 model \cite{vgg8_ref} used for image classification from the CIFAR-10 dataset \cite{vgg8_code}.}
	\label{vgg8}
\end{figure}
\par
The hardware implementation for on-chip learning is shown in Fig. \ref{dnn_arch}. It consists of crossbar arrays integrated with peripheral read/write circuits, analog-to-digital converters (ADCs), multiplexers, and adders, forming a transposable synaptic array (TSA). Multiple TSAs are interconnected using H-routing with embedded buffers to construct processing elements (PEs), which are then organized into tiles. The high-level architecture comprises multiple tiles, each incorporating dedicated units for weight gradient computation, global buffering, accumulation, activation, and pooling operations.
Weight updates are performed sequentially in a row-by-row manner, whereas inference is executed in parallel by activating all columns simultaneously. 
Write and read lines regulate access transistors, enabling selective read and write operations for individual synaptic devices. To optimize energy and area efficiency, the column multiplexer employs column sharing, with one ADC shared across eight columns. Along each column, the output vectors are initially generated as analog partial current sums, which are then digitized by the ADCs. The final summation of multi-state weights and input multiplications is carried out using shift-and-add digital processing modules.
\begin{figure*}
	\centering
	\begin{subfigure}{0.33\linewidth}
		\includegraphics[width=\linewidth, keepaspectratio]{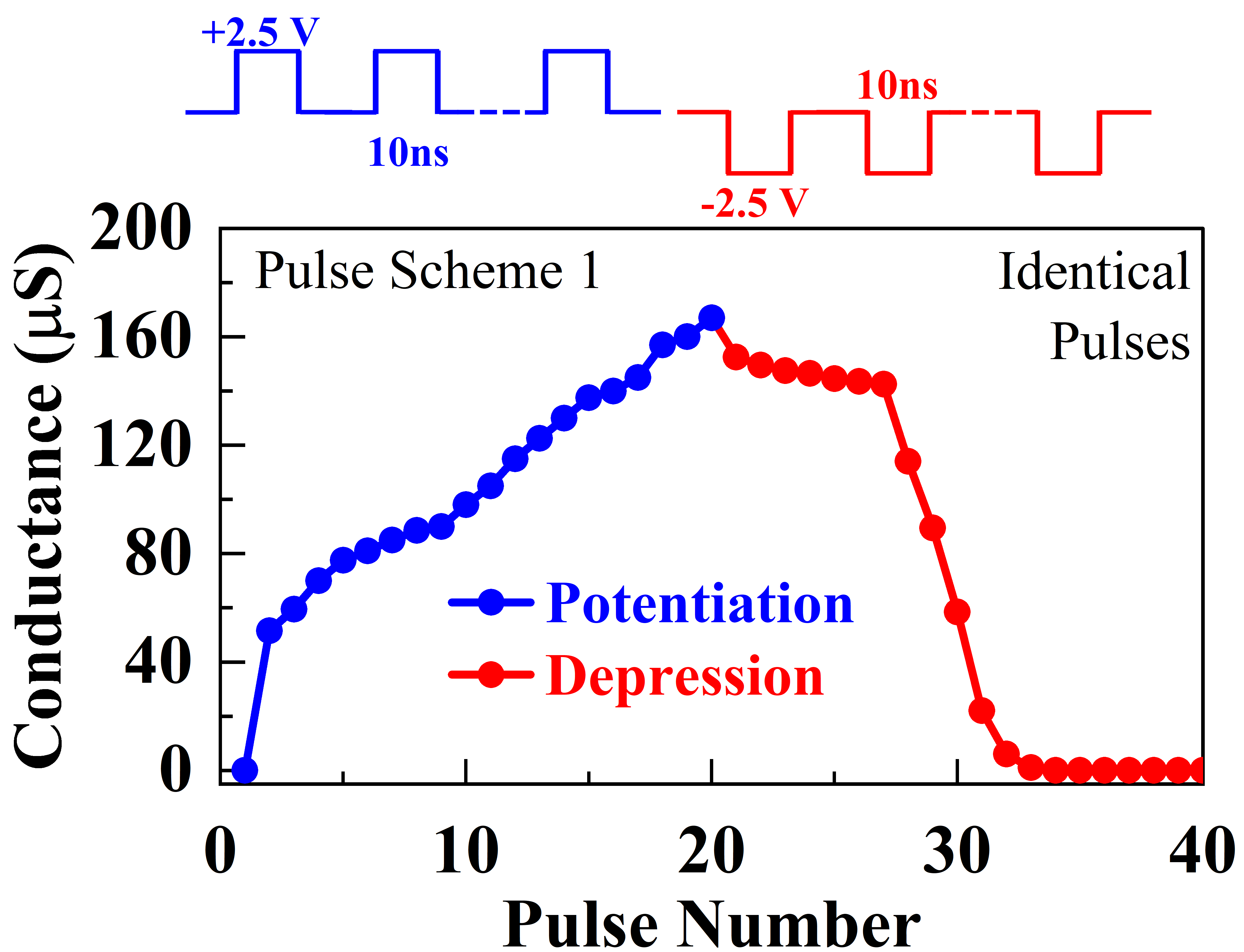}
		\caption{}
		\label{pulse1}
	\end{subfigure}\hfil
	\begin{subfigure}{0.33\linewidth}
		\includegraphics[width=\linewidth, keepaspectratio]{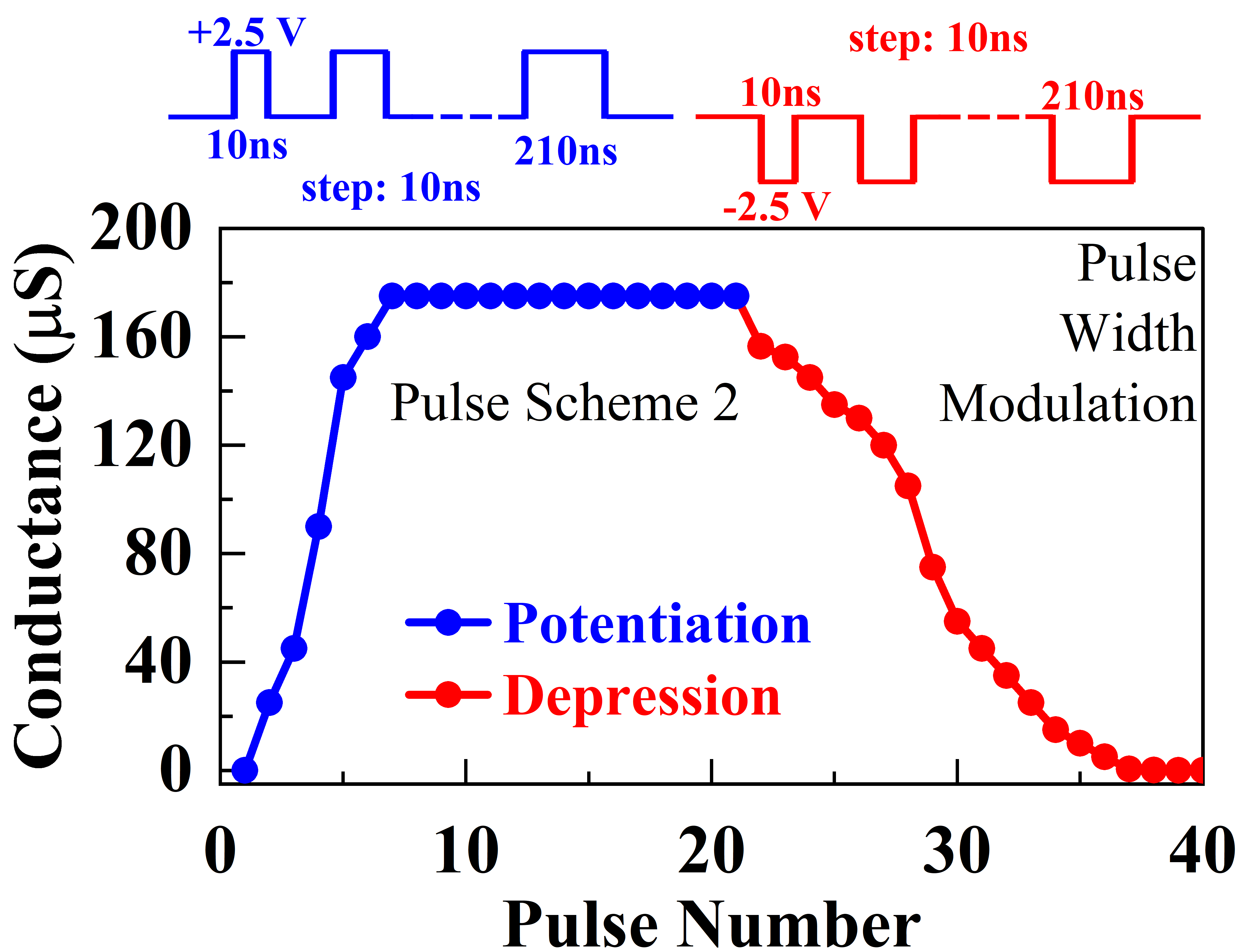}
		\caption{}
		\label{pulse2}
	\end{subfigure}\hfil
	\begin{subfigure}{0.33\linewidth}
		\includegraphics[width=\linewidth, keepaspectratio]{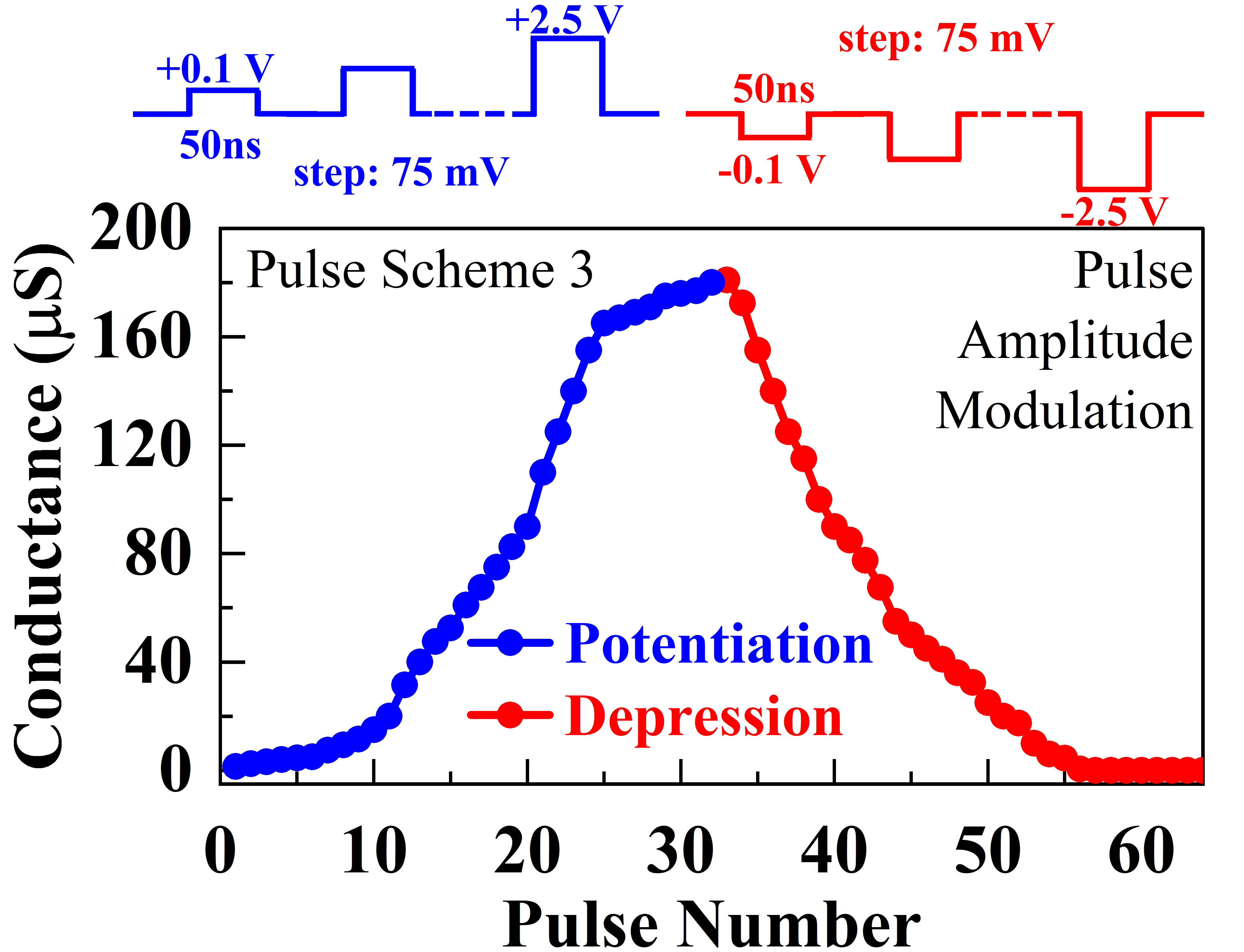}
		\caption{}
		\label{pulse3}
	\end{subfigure}\hfil
	\caption{Simulated response of an ULTRA\textbf{RAM} cell to (a) identical pulses (same magnitude and pulse width), (b) variable pulse width for a fixed voltage magnitude, and (c) variable amplitude for a fixed pulse width. The number of accessible partial states is maximized when using a variable amplitude pulse scheme ($\sim 32$ states for LTP and LTD).}
	\label{fig_pulses}
\end{figure*}
\par
The pseudo crossbar array consists of an access transistor along with each memory cell. The access transistor ensures that only the selected rows are programmed during the row-by-row weight update, preventing unintended programming of other rows. ULTRA\textbf{RAM} cells operate as a three-terminal device (assuming back-gate grounded) and requires two separate input signals: one for activating the word lines (WLs) and another for applying read voltages to the read select (RS). The RS facilitates the retrieval of input vectors, as shown in Fig. \ref{dnn_arch}.
\par
The VGG-8 architecture is utilized for classifying 32$\times$32 color images from the CIFAR-10 dataset, as illustrated in Fig. \ref{vgg8} \cite{vgg8_ref, vgg8_code}. This network comprises six convolutional layers ($C_1$--$C_6$) followed by two fully connected layers ($FC_1$ and $FC_2$) for image classification. Max-pooling layers with a 2$\times$2 kernel are applied after each convolutional layer to downsample feature maps.
%\par
To process an image (inferencing), input voltages corresponding to 1024 extracted features from the 32$\times$32 image are applied to the crossbar array. The read voltages, representing the element-wise product of input values and synaptic weights, accumulate based on Kirchhoff's law and are subsequently fed into the activation function circuit at each output node. This enables efficient matrix-vector multiplication directly within the crossbar array.
\par
For network training, the stochastic gradient descent algorithm is used to determine the weight updates at each output node, facilitated by a dedicated weight update circuit. The computed weight changes are then multiplied by the corresponding inputs using a multiplier circuit. The resulting voltages from the multiplier serve as a programming voltage for the ULTRA\textbf{RAM} synapse, adjusting its conductance to reflect the updated weight values.
%
% ==================================================
\section{Non-ideal Synaptic Device Properties}
The conductance of synaptic devices can be adjusted by applying positive or negative programming voltage pulses, corresponding to weight increment and decrement, respectively. Ideally, a synaptic device exhibits a linear weight update response to uniform programming voltage pulses.
However, practical devices might deviate from this ideal behavior, displaying "non-ideal" characteristics such as nonlinear and fluctuating weight updates. This can restrict precision and lead to a finite ON/OFF ratio.
%\par
We have analyzed the long-term potentiation (LTP) and long-term depression (LTD) behavior of ULTRA\textbf{RAM} devices under different pulse schemes.
Fig. \ref{pulse1} shows the Scheme 1 with identical pulses. Each programming pulse has the same amplitude and duration for both potentiation and depression. In Scheme 2, the applied pulse width is varied gradually, keeping magnitude constant, to control the weight update, as shown in Fig. \ref{pulse2}. Lastly, in Scheme 3, we have applied a fixed time period pulse ($50ns$) width varying pulse magnitude from $\pm 0.1V$ to $\pm 2.5V$, as shown in Fig. \ref{pulse3}. The Scheme 3 shows the linear weight update in both potentiation and depression compared to other two schemes. In addition, it provides the maximum number of accessible partial states compared to the other schemes.
The conductance change with a number of pulses (P) is fitted and non-linearity in LTP and LTD are extracted by the method in the DNN+NeuroSim Framework \cite{ch8dnn4} as follows:
\begin{equation}
	G_{LTP} = B \left( 1 - exp\left(- \frac{P}{\alpha_p} \right)  \right) + G_{min}
	\label{gltp}
\end{equation} 
\begin{equation}
	G_{LTD} = -B \left( 1 - exp\left(\frac{P - P_{max}}{\alpha_d} \right)  \right) + G_{max}
	\label{gltd}
\end{equation} 
\begin{equation}
	B = (G_{max} - G_{min}) \text{\huge/} \left( 1 - exp\left(\frac{-P_{max}}{\alpha_{p,d}} \right)  \right)
	\label{eqn_b}
\end{equation} 
where, $G_{LTP}$ and $G_{LTD}$ are the conductance for LTP and LTD, respectively. $G_{max}$, $G_{min}$ and $P_{max}$ are the maximum conductance, minimum conductance and the maximum pulse number required to switch the device between the minimum and maximum conductance states, respectively. $\alpha_{p,d}$ is the parameter that controls the nonlinear behavior of weight update, and $B$ is simply a function of $\alpha_{p,d}$ that fits the functions within the range of $G_{max}$, $G_{min}$ and $P_{max}$.
Scheme 3 exhibits the greatest number of states with symmetric response due to optimal sampling of charge storage in the FG through TBRT. Therefore, we have considered this scheme for on-chip training using ULTRA\textbf{RAM} cells.
%
% ======= Latency and Energy UR.Si ===============
\begin{figure*}[t!]
	\centering
	\includegraphics[width=0.9\linewidth]{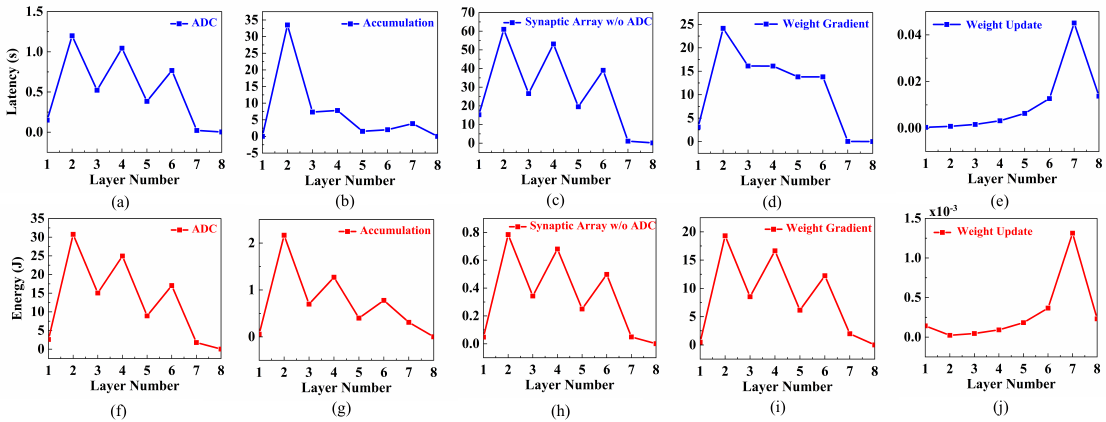}
	\caption{(a)-(e) Peak latency and (f)-(j) energy across all the layers in VGG-8 for various CNN modules/operations (ADC, accumulation, synaptic array, weight gradient calculation, and weight update) in one epoch. The data shown is from the 256th epoch of 2-bit ULTRA\textbf{RAM}-based CIM architecture.}
	\label{fig_latency}
\end{figure*}
\section{Performance of CNN}
%First, we have used the experimentally demonstrated ULTRA\textbf{RAM} cells to get the actual on-chip performance. As the currently fabricated cells have long channel lengths ($\sim$$10\mu m$), we have compared them with conventional SRAM-based synapses for better estimation of performance matrices. In addition, current cells have a limited ON/OFF ratio to get the multi-states which can be seen in Fig. \ref{fig_idvg}. Second, we have projected the on-chip performance with scaled-down simulated devices that match the current technology nodes of other emerging memory devices. 
The performance of CNNs was evaluated using experimentally demonstrated long-channel-length ULTRA\textbf{RAM} cells, and projected the performance with simulated devices at scaled technology nodes.
A physics-based model has been used to investigate the experimental and theoretical response of ULTRA\textbf{RAM} cells for various pulse schemes. A detailed description of the model can be found in \cite{mypaper_uredtm, mypaper_urdrc}. Then, a synaptic crossbar array of size 128$\times$128 has been considered for simulations using the DNN+NeuroSIM simulator for each layer separately.
%
%
%\subsection{Comparison of Actual Devices}
\subsection{Long-channel Devices}
%ULTRA\textbf{RAM} cells have been fabricated on GaAs and Si substrates with 10 $\mu m$ of channel lengths \cite{ch7ur1.ted, ch7ur2.aem}. The program/erase operation of GaAs substrate devices has been performed using the $\pm 2.5V$ control gate pulses magnitude with 500$\mu s$ pulse widths \cite{ch7ur1.ted}. The devices on Si substrate uses the same program/erase voltages with 10$ms$ of pulse width \cite{ch7ur2.aem}. Additionally, we have considered an ideal ULTRA\textbf{RAM} device with similar dimensions and improved output characteristics compared to the experimental one. This can help us to predict the on-chip performance with improved device characteristics and can be used as the design guidelines for present ULTRA\textbf{RAM} cells. 
We have considered two types of long-channel device for on-chip performance simulations. (1) ULTRA\textbf{RAM} cells fabricated on GaAs and Si substrates with 10 $\mu m$ of channel lengths \cite{ch7ur1.ted, ch7ur2.aem}. These devices exhibit a limited current ratio, which restricts the number of achievable conductance states (2-bit), as shown in Fig. \ref{fig_idvg}. Nevertheless, appropriate device design and optimization can significantly improve their output characteristics upto 5-bit/cell with similar device dimensions \cite{myjap_UR}. (2) We have also considered these improved characteristics ULTRA\textbf{RAM} cells (5-bit) with similar device dimensions and used to predict the potential on-chip performance with optimized properties. This can serve as design guidelines for advancing present ULTRA\textbf{RAM} technology.
%
% ======= Accuracy vs. Epochs ===============
\begin{figure}[h!]
	\centering
	\includegraphics[width=0.8\columnwidth]{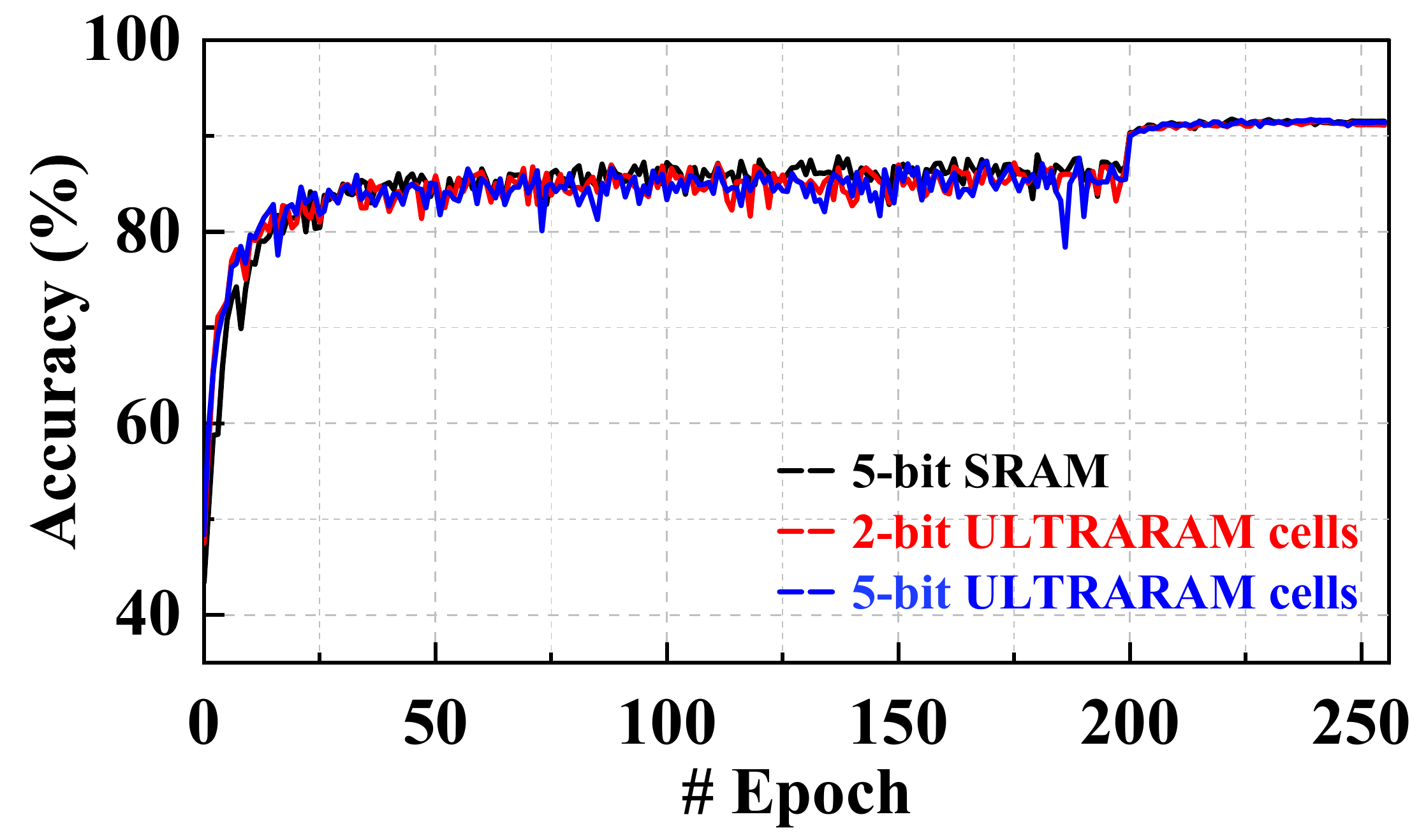}
	\caption{Accuracy achieved for 5-bit SRAM, 2-bit experimentally demonstrated ULTRA\textbf{RAM}, and 5-bit simulated ULTRA\textbf{RAM} device precision in 256 epochs.}
	\label{fig_acc_epoch}
\end{figure}
\par
The full set of performance metrics is obtained over 256 epochs. Fig. \ref{fig_latency} shows the latency and energy consumption for each layer of various CNN modules and operations. This includes the ADC, accumulation, synaptic array, weight gradient computation, and weight update. The overall energy and latency are primarily influenced by four key processes: feedforward, error computation, gradient computation, and weight update. Among these, weight gradient computation significantly impacts both energy and latency due to the frequent read and write operations required for activation functions and error processing.
%
%
% ========== Latency and Energy UR.Si ================
%\begin{figure}
%	\centering
%	\begin{subfigure}{0.2\columnwidth}
	%		\includegraphics[width=\linewidth, keepaspectratio]{./images/ADC_latency.pdf}
	%		\caption{}
	%		\label{adc_lat}
	%	\end{subfigure}	\hfil
%	\begin{subfigure}{0.2\linewidth}
	%		\includegraphics[width=\textwidth, keepaspectratio]{./images/Accumulation_letency.pdf} 
	%		\caption{}
	%		\label{acc_lat}
	%	\end{subfigure}\hfil
%	\begin{subfigure}{0.2\linewidth}
	%		\includegraphics[width=\linewidth, keepaspectratio]{./images/synaptic_array_latency.pdf}
	%		\caption{}
	%		\label{syn_lat}
	%	\end{subfigure}\hfil
%	\begin{subfigure}{0.2\linewidth}
	%		\includegraphics[width=\linewidth, keepaspectratio]{./images/WG_latency.pdf}
	%		\caption{}
	%		\label{wg_lat}
	%	\end{subfigure}\hfil
%	\begin{subfigure}{0.2\linewidth}
	%		\includegraphics[width=\linewidth, keepaspectratio]{./images/WU_latency.pdf}
	%		\caption{}
	%		\label{wu_lat}
	%	\end{subfigure}
%	%
%\caption{}
%\label{fig_latency}
%\end{figure}
%
%
\par
To assess the influence of an ULTRA\textbf{RAM} synapse on a CNN's performance, the proposed 2-bit and 5-bit ULTRA\textbf{RAM}-based CNNs were evaluated in comparison to a 5-bit SRAM-based CNN using the same simulation framework. Fig. \ref{fig_acc_epoch} shows the relationship between the number of training epochs and the accuracy of 5-bit SRAM and two different ULTRA\textbf{RAM} cells implemented with 2-bit and 5-bit weight precision.
It is observed that the ULTRA\textbf{RAM}-based neural network demonstrates accuracy comparable to that of a 5-bit SRAM-based design. However, the 2-bit ULTRA\textbf{RAM}-based CNN exhibits superior efficiency, being 1.8$\times$ more area-efficient and 1.52$\times$ more energy-efficient. However, it loses in terms of latency and can be seen in Table \ref{ch8_tab_exp}. For a fair comparison, we have compared 5-bit SRAM with a 5-bit simulated ULTRA\textbf{RAM}-based CNN. This results in improvement in area, energy, and latency by 3.38$\times$, 2.06$\times$, and 1.25$\times$, respectively, compared to 5-bit SRAM-based CNN without affecting the accuracy and can be seen in Fig. \ref{fig_acc_epoch}.
%
%
%============= Tab1 Benchmark with exp UR ===============
\begin{table*}
	\caption{\\Benchmark results of CIM accelerators training on VGG-8 for CIFAR-10, based on SRAM and Long-channel ULTRA\textbf{RAM} synaptic cells with 256 epochs.}
	%	\scriptsize
	\setlength{\tabcolsep}{8pt}
	\begin{center}
		\renewcommand{\arraystretch}{1.2}
		\begin{tabular}{c  c  c  c c }
			%\begin{tabular}{p{55pt} p{125pt} p{100pt} p{100pt} p{100pt}}
			\hline
			\textbf{Technology Node} & \multicolumn{4}{c}{\textbf{130 nm}}\\
			\hline 
			\multirow{2}{*}{\textbf{Device}} & \multirow{2}{*}{\textbf{SRAM}}   & \textbf{ULTRA\textbf{RAM}} & \textbf{ULTRA\textbf{RAM}} & \textbf{ULTRA\textbf{RAM}}   \\
			&  & \textbf{(GaAs Subs.) \cite{ch7ur1.ted} }   & \textbf{(Si Subs.) \cite{ch7ur2.aem}}  & \textbf{(Optimized)*}   \\
			\hline
			\textbf{\# Conductance States} & 32 & 4 & 4 & 32 \\
			%			\textbf{ADC Precision} & 6 & 6 & 6 & 6 \\
			\textbf{Cell Precision} & 1-bit & 2-bit & 2-bit & 5-bit \\
			\hline
			\textbf{R$_{ON}$ [$\Omega$]} & -- & 0.6$K$ & 0.33$K$ & 5$K$ \\
			\textbf{ON/OFF Ratio} & -- & 2 & 2 & 10 \\
			\textbf{C2C Variation} & -- & $<$0.5\% & $<$0.5\% & 3\% \\
			\textbf{Write Pulse Voltage [V]} & -- & $\pm$2.5 & $\pm$2.5 & $\pm$2.5  \\
			\textbf{Write Pulse Width} & -- & 500 $\mu s$ & 10 $ms$ & 100 $ns$ \\
			%			\textbf{Read Pulse Voltage [V]} & -- & 0.5 & 0.2 & 0.2 \\
			%			\textbf{Read Pulse Width} & -- & 500 $\mu s$ & 10 $ms$ & 50 $ns$ \\
			\hline
			\textbf{Area [$mm^2$]} & 6295.3 & 3491 & 3576 & 1862 \\
			\textbf{Memory Utilization (\%)} & 94.62 & 88.59 & 88.59 & 88.59 \\
			\hline
			\textbf{Training Accuracy} & 91.7 & 91.52 & 91.68 & 91.69 \\
			\hline
			\textbf{Training Latency (s) / Epoch} & 453.2 & 490.4 & 588 & 362.12 \\
			\textbf{Training Dynamic Energy (J) / Epoch} & 358.4 & 235.43 & 267 & 173.6 \\
			%			\textbf{Training Peak Latency (s) / Epoch} & 184.6 & 266.13 & 363 & 161.9 \\
			%			\textbf{Training Peak Dynamic Energy (J) / Epoch} & 258 & 142.4 & 174 & 83.9 \\
			\hline
			\textbf{Training Throughput (TOPS)} & 0.406 & 0.376 & 0.31 & 0.50 \\
			\textbf{Training Energy Efficiency (TOPS/W)} & 0.508 & 0.781 & 0.68 & 1.06 \\
			%			\textbf{Training Peak Throughput (TOPS)} & 0.998 & 0.6936 & 0.5 & 1.13 \\
			%			\textbf{Training Peak Energy Efficiency (TOPS/W)} & 0.714 & 1.295 & 1.05 & 2.19 \\
			\hline
			
			\multicolumn{5}{p{420pt}}{$^*$Projected performance from long channel devices with optimized characteristics.}\\
		\end{tabular}
		\label{ch8_tab_exp}
	\end{center}
\end{table*}
%
% ======= Latency and Energy UR 32nm ===============
\begin{figure*}[b!]
	\centering
	\includegraphics[width=0.9\linewidth]{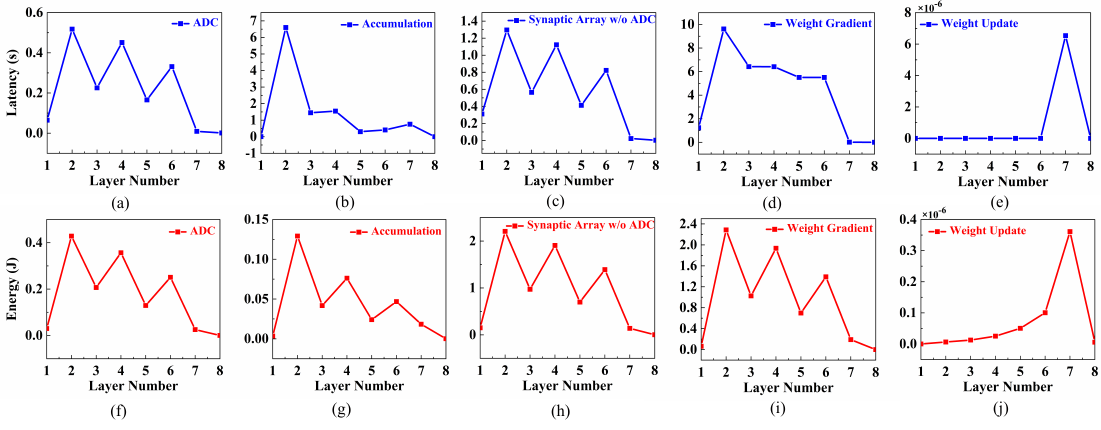}
	\caption{(a)-(e) Peak latency and (f)-(j) energy across all the layers in VGG-8 for various CNN modules/operations (ADC, accumulation, synaptic array, weight gradient calculation, and weight update) in one epoch. The data shown is from the 256th epoch of simulated 5-bit ULTRA\textbf{RAM}-based CIM architecture.}
	\label{fig_latency_32nm}
\end{figure*}
\par
Finally, we have evaluated the performance of CIM accelerators for VGG-8 training on the CIFAR-10 dataset \cite{vgg8_ref, vgg8_code}, utilizing ULTRA\textbf{RAM} and SRAM-based accelerators. Due to the longer channel lengths ($>$$10 \mu m$) of experimentally demonstrated ULTRA\textbf{RAM} cells, we have assumed 130 nm technology node for evaluating the on-chip performance. Table \ref{ch8_tab_exp} shows the benchmark results of CIM accelerators based on SRAM and ULTRA\textbf{RAM} synaptic cells with 256 epochs.
The on-chip 5-bit SRAM-based CMOS implementation provides the same training accuracy but requires a significantly larger chip area overhead relative to 2-bit ULTRA\textbf{RAM} non-volatile memory cells.
Additionally, the 2-bit ULTRA\textbf{RAM} synapses exhibit a comparable energy, latency and TOPS advantage compared to 5-bit SRAM-based synapses. These performance parameters can be further improved by using a optimized 5-bit ULTRA\textbf{RAM}-based synapses, as projected in Table \ref{ch8_tab_exp}.
\subsection{Projection with Scaled Devices}
We have also simulated the ULTRA\textbf{RAM} cells with scaled-down channel lengths ($\sim$100 $nm$) considering the same TBRT stack replacing the gate oxide.
Now, we have compared this with other analog emerging memory devices at 32-nm technology nodes.  %
\par
\begin{figure}
	\centering
	\includegraphics[width=0.8\columnwidth]{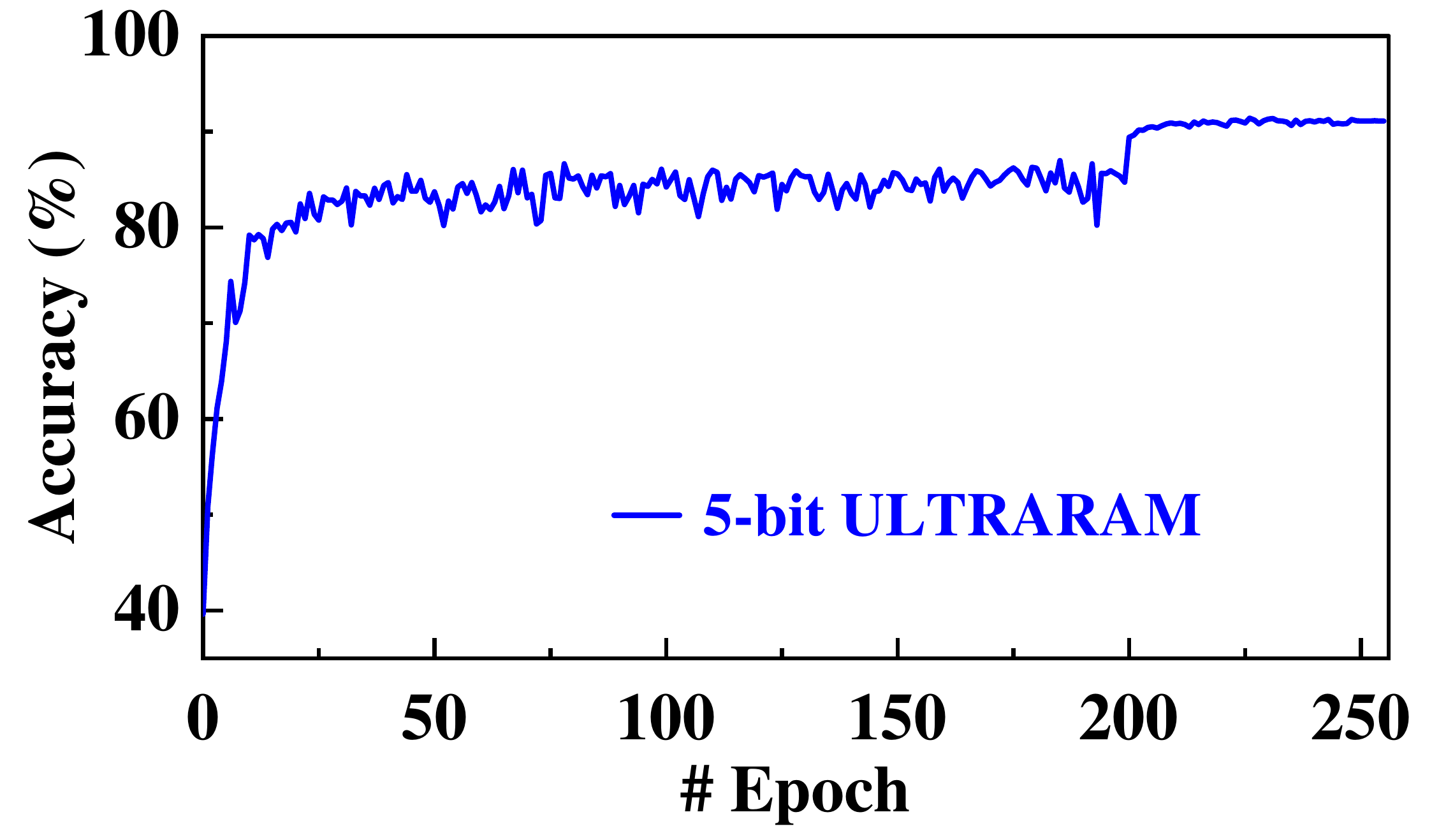}
	\caption{Accuracy achieved in 256 epochs of 5-bit ULTRA\textbf{RAM}-based CIM architecture at 32-nm technology node.}
	\label{fig_acc_epoch_32nm}
\end{figure}
\begin{figure}
	\centering
	\includegraphics[width=0.8\columnwidth]{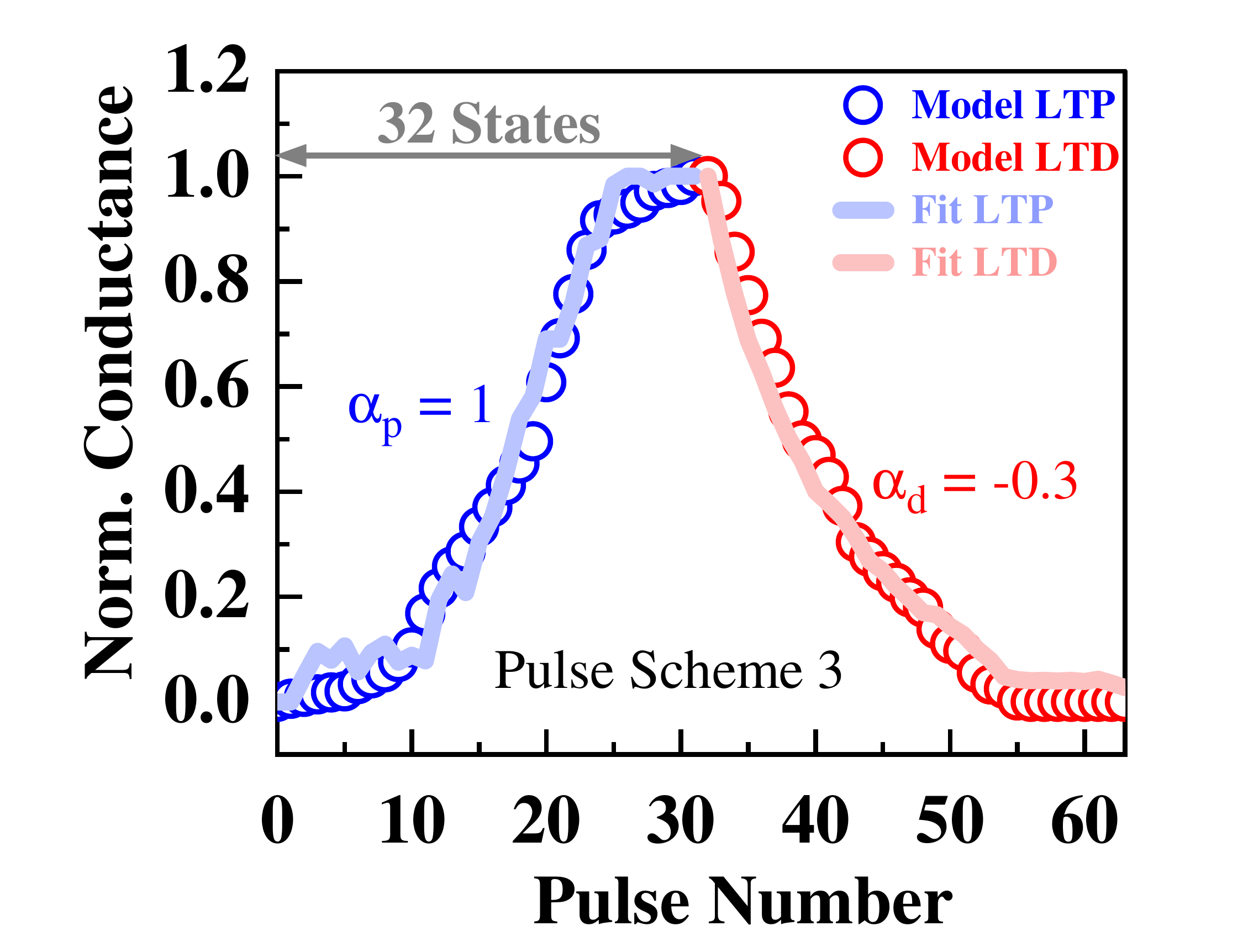}
	\caption{Normalized simulated response of a 32-nm node ULTRA\textbf{RAM} cell using pulse Scheme 3 (varying magnitude with a fixed pulse width). The corresponding non-linearity ($\alpha_{p/d}$) has been extracted using the equations \eqref{gltp} and \eqref{gltd}.}
	\label{norm_pulse3}
\end{figure}

%
%
%
%================ Tab3 Benchmark =============================
\begin{table*}
	\caption{\\Benchmark results of CIM accelerators training on VGG-8 for CIFAR-10, based on SRAM, reported analog synaptic devices, and ULTRA\textbf{RAM} synaptic cells with 256 epochs.}
	%	\scriptsize
	\setlength{\tabcolsep}{2pt}
	\begin{center}
		\renewcommand{\arraystretch}{1.2}
		\begin{tabular}{c  c  c  c c c  c c }
			%\begin{tabular}{p{55pt} p{125pt} p{100pt} p{100pt} p{100pt}}
			\hline
			\textbf{Technology Node} & \multicolumn{7}{c}{\textbf{32 nm}} \\
			\hline 
			\multirow{3}{*}{\textbf{Device}} & \multirow{3}{*}{\textbf{SRAM}}  & \textbf{Memristor} & \textbf{RRAM} & \textbf{RRAM}  & \textbf{EpiRAM}  & \textbf{FeFET}  & \textbf{ULTRA\textbf{RAM}*}    \\
			&  & \textbf{\cite{ch8memristor1} }   & \textbf{ (PCMO)}  & \textbf{(AlO$_x$/HfO$_2$)}  & \textbf{\cite{ch8epiram}}   & \textbf{\cite{ch5neuro1}}   & \textbf{(This Work)}  \\
			&  &  & \textbf{\cite{ch8_benchrram1}}  & \textbf{ \cite{ch8rram4}}  &    &   &     \\
			\hline
			\textbf{\# Conductance States} & -- & 97 & 50 & 40 & 64  & 32  & 32   \\
			%			\textbf{ADC Precision} & 4 & 6 & 6 & 6  & 6  & 6  & 6  \\
			\textbf{Cell Precision} & 1 & 6 & 5 & 5  & 6  & 5  & 5  \\
			\hline
			\textbf{R$_{ON}$ [$\Omega$]} & -- & 26$M$ & 23$M$ & 16.9$K$  & 81$K$  & 240$K$  & 5$K$  \\
			\textbf{ON/OFF Ratio} & -- & 12.5 & 6.84 & 4.43  & 50.2  & 10  & 10  \\
			\textbf{C2C Variation (\%)} & -- & 3.5 & $<$1 & 5 & 2  & $<$0.5  & 3  \\
			\textbf{Write Pulse Voltage [V]} & -- & $\pm$3 & $\pm$2 & $\pm$1  & $\pm$5  & $\pm$4  & $\pm$2.5  \\
			\textbf{Write Pulse Width} & -- & 300 $\mu s$ & 1 $ms$ & 100 $\mu s$  & 5 $\mu s$  & 50 $ns$  & 50 $ns$   \\
			\hline
			\textbf{Area [$mm^2$]} & 138.95 & 48.29 & 48.29 & 49.88  & 48.59  & 95.21  & 101.48   \\
			\textbf{Memory Utilization (\%)} & 94.62 & 88.59 & 88.59 & 88.59  & 88.59  & 88.59  & 88.59    \\
			\hline
			\textbf{Training Accuracy (\%)} & 91 & 49 & 56 & 37 & 85  & 91.12  & 91.28   \\
			\hline
			\textbf{Training Latency (s) / Epoch} & 235.75 & 1241.63 & 5795.79 & 611  & 193.94  & 121.66  & 125.9   \\
			\textbf{Training Dynamic Energy (J) / Epoch} & 95.37 & 92.12 & 92.15 & 93.13 & 92.28  & 87.18  & 86.68   \\
			%			\textbf{Training Peak Latency (s) / Epoch} & 55.33 & 1116.53 & 5670.69 & 485.89 & 68.83  & 48.61  & 52.10   \\
			%			\textbf{Training Peak Dynamic Energy (J) / Epoch} & 15.42 & 9 & 9 & 10.70 & 9.19  & 9.72  & 9.42   \\
			\hline
			\textbf{Training Throughput (TOPS)} & 0.78 & 0.14 & 0.003 & 0.30 & 0.95  & 1.51  & 1.46   \\
			\textbf{Training Energy Efficiency (TOPS/W)} & 1.94 & 2 & 2 & 1.98  & 2  & 2.11  & 2.12  \\
			%			\textbf{Training Peak Throughput (TOPS)} & 3.34 & 0.16 & 0.03 &  0.38 & 2.68  & 3.79  & 3.53   \\
			%			\textbf{Training Peak Energy Efficiency (TOPS/W)} & 11.98 & 20.54 & 20.50 & 17.27 & 20.11  & 18.95  & 19.55   \\
			\hline
			
			\multicolumn{8}{p{380pt}}{$^*$Projected performance with 32 nm technology node scaled device parameters simulated with model.}\\
		\end{tabular}
		\label{ch8_tab_projection}
	\end{center}
\end{table*}
Fig. \ref{fig_latency_32nm} shows the latency and energy consumption for each layer of various CNN modules and operations considering the 5-bit ULTRA\textbf{RAM}-based synapse. This shows that the latency and energy consumption can be significantly reduced with the scaled ULTRA\textbf{RAM} cells as compared to experimentally demonstrated cells [Fig. \ref{fig_latency}]. In addition, the training accuracy is comparable to the existing ULTRA\textbf{RAM} cells with 3\% of cycle-to-cycle (C2C) variations, as shown in Fig. \ref{fig_acc_epoch_32nm}.
We have used the pulse Scheme 3 (pulse amplitude modulation) to plot the conductance change with the number of pulses (P) and non-linearity in LTP and LTD using the equations \eqref{gltp} and \eqref{gltd}, as shown in Fig. \ref{norm_pulse3}.
%
% ======= Accuracy vs. Epochs UR 32nm===============
The 5-bit ULTRA\textbf{RAM}-based CNN exhibits better efficiency, being 1.36$\times$ more area-efficient, 1.1$\times$ more energy-efficient, and 1.87$\times$ faster in terms of latency compared to 32-nm node SRAM-based CNN.
\par
Finally, we have benchmarked the performance of CIM accelerators utilizing various analog synaptic devices, including memristor, RRAM, EpiRAM, and FeFET, with ULTRA\textbf{RAM}-based synapse at 32-nm technology node, as shown in Table \ref{ch8_tab_projection}.
It is observed that the ULTRA\textbf{RAM}-based synapse can provide better performance in terms of throughput, area, latency, and energy compared to SRAM. Performance is comparable to FeFET devices, suggesting that scaled ULTRA\textbf{RAM}-based CNNSs can be used as a artificial synapes for DNN acceleration.
%
%
%
% ======= Norm. pulse 3 ===============
\par
ULTRA\textbf{RAM} memory shows promise as a synaptic cell for DNN acceleration.
Based on the hardware performance results presented in Tables \ref{ch8_tab_exp} and \ref{ch8_tab_projection}, the following observations can be made:
(i) Optimizing on-state resistance ($R_{ON}$) is critical for minimizing voltage drops; however, scaling transistors in 1T1R architectures or peripheral multiplexers increases area overhead and parasitic capacitance, adversely impacting latency and throughput. (ii) Write pulse durations below a microsecond remain unaffected due to batch-wise amortization. (iii) Maintaining cycle-to-cycle variation below 1\% is essential to ensure stable in-situ training, as higher variations can disrupt model convergence. (iv) While SRAM-based architectures encounter leakage and area constraints at larger technology nodes, parallel-read SRAM designs at advanced nodes offer superior energy efficiency and throughput.
%
%\par
%The performance can be further improved if the following remaining challenges can be addressed: (i) the improvement in ON/OFF contrast ratio to get the more number of intermediate states, (ii) the scalability towards sub-20nm technology nodes, and (iii) the program/erase voltage ($\sim \pm2.5 V$) is remains too high to align with the power supply of logic transistors in scaled technology nodes, which requires a level shifter to convert the voltages between the ULTRA\textbf{RAM} array and the peripheral circuitry.
%
\section{Conclusion}
In this work, we have presented on-chip training and inference of a neural network using ULTRA\textbf{RAM} memory device-based synaptic arrays. The longer channel 2-bit ULTRA\textbf{RAM}-based CNN exhibits superior efficiency, being 1.8$\times$ more area-efficient and 1.52$\times$ more energy-efficient. Additionally, the performance projection has been demonstrated with the simulated ULTRA\textbf{RAM} cells scaled down to advanced technology nodes (32-nm). This results superior performance than SRAM- and several emerging memory technologies-based CNN implementations, while maintaining performance levels comparable to FeFET-based designs with respect to critical system metrics such as area, latency, energy consumption, and throughput.

\section*{\textbf{Acknowledgments}}
This work was supported in parts by the Quinas Technology Limited, Lancaster, United Kingdom; Indian Institute of Technology Roorkee, India, and in part by the Prime Minister's Research Fellowship, Ministry of Education, Government of India under Grant PM-31-22-773-414.

\section*{\textbf{Conflict of Interest}}
The authors have no conflicts to disclose.

\section*{\textbf{Data Availability Statement}}
The data that support the findings of this study are available within the article.

\appendices

\bibliographystyle{IEEEtran}
\bibliography{Bibliography}

\begin{thebibliography}{10}
\providecommand{\url}[1]{#1}
\csname url@samestyle\endcsname
\providecommand{\newblock}{\relax}
\providecommand{\bibinfo}[2]{#2}
\providecommand{\BIBentrySTDinterwordspacing}{\spaceskip=0pt\relax}
\providecommand{\BIBentryALTinterwordstretchfactor}{4}
\providecommand{\BIBentryALTinterwordspacing}{\spaceskip=\fontdimen2\font plus
\BIBentryALTinterwordstretchfactor\fontdimen3\font minus
  \fontdimen4\font\relax}
\providecommand{\BIBforeignlanguage}[2]{{%
\expandafter\ifx\csname l@#1\endcsname\relax
\typeout{** WARNING: IEEEtran.bst: No hyphenation pattern has been}%
\typeout{** loaded for the language `#1'. Using the pattern for}%
\typeout{** the default language instead.}%
\else
\language=\csname l@#1\endcsname
\fi
#2}}
\providecommand{\BIBdecl}{\relax}
\BIBdecl

\bibitem{ch8dnn1}
J.~Schmidhuber, ``Deep learning in neural networks: An overview,'' \emph{Neural
  networks}, vol.~61, pp. 85--117, 2015.

\bibitem{ch8dnn3}
N.~Rusk, ``Deep learning,'' \emph{Nature Methods}, vol.~13, no.~1, pp. 35--35,
  2016.

\bibitem{ch1r2}
H.-S.~P. Wong and S.~Salahuddin, ``Memory leads the way to better computing,''
  \emph{Nature nanotechnology}, vol.~10, no.~3, pp. 191--194, 2015.

\bibitem{ch8intro3}
C.-J. Jhang, C.-X. Xue, J.-M. Hung, F.-C. Chang, and M.-F. Chang, ``Challenges
  and trends of sram-based computing-in-memory for ai edge devices,''
  \emph{IEEE Transactions on Circuits and Systems I: Regular Papers}, vol.~68,
  no.~5, pp. 1773--1786, 2021.

\bibitem{ch8intro2}
K.~Yu, S.~Kim, and J.~R. Choi, ``Trends and challenges in computing-in-memory
  for neural network model: A review from device design to application-side
  optimization,'' \emph{IEEE Access}, 2024.

\bibitem{ch8intro4}
S.~Mittal, G.~Verma, B.~Kaushik, and F.~A. Khanday, ``A survey of sram-based
  in-memory computing techniques and applications,'' \emph{Journal of Systems
  Architecture}, vol. 119, p. 102276, 2021.

\bibitem{ch8dram1}
F.~Gao, G.~Tziantzioulis, and D.~Wentzlaff, ``Computedram: In-memory compute
  using off-the-shelf drams,'' in \emph{Proceedings of the 52nd annual IEEE/ACM
  international symposium on microarchitecture}, 2019, pp. 100--113.

\bibitem{ch8dram2}
S.~Khoram, Y.~Zha, J.~Zhang, and J.~Li, ``Challenges and opportunities: From
  near-memory computing to in-memory computing,'' in \emph{Proceedings of the
  2017 ACM on International Symposium on Physical Design}, 2017, pp. 43--46.

\bibitem{ch5app1}
S.~Kim and H.-J. Yoo, ``An overview of computing-in-memory circuits with dram
  and nvm,'' \emph{IEEE Transactions on Circuits and Systems II: Express
  Briefs}, vol.~71, no.~3, pp. 1626--1631, 2024.

\bibitem{ch8neuro1}
S.~Dutta, H.~Ye, W.~Chakraborty, Y.-C. Luo, M.~San~Jose, B.~Grisafe, A.~Khanna,
  I.~Lightcap, S.~Shinde, S.~Yu \emph{et~al.}, ``Monolithic 3d integration of
  high endurance multi-bit ferroelectric fet for accelerating
  compute-in-memory,'' in \emph{2020 IEEE International Electron Devices
  Meeting (IEDM)}.\hskip 1em plus 0.5em minus 0.4em\relax IEEE, 2020, pp.
  36--4.

\bibitem{ch8rram1}
S.~Yin, Y.~Kim, X.~Han, H.~Barnaby, S.~Yu, Y.~Luo, W.~He, X.~Sun, J.-J. Kim,
  and J.-s. Seo, ``Monolithically integrated rram-and cmos-based in-memory
  computing optimizations for efficient deep learning,'' \emph{IEEE Micro},
  vol.~39, no.~6, pp. 54--63, 2019.

\bibitem{ch8rram3}
G.~Pedretti and D.~Ielmini, ``In-memory computing with resistive memory
  circuits: Status and outlook,'' \emph{Electronics}, vol.~10, no.~9, p. 1063,
  2021.

\bibitem{ch8pcram1}
Q.~Wang, G.~Niu, W.~Ren, R.~Wang, X.~Chen, X.~Li, Z.-G. Ye, Y.-H. Xie, S.~Song,
  and Z.~Song, ``Phase change random access memory for neuro-inspired
  computing,'' \emph{Advanced Electronic Materials}, vol.~7, no.~6, p. 2001241,
  2021.

\bibitem{ch5neuro1}
M.~Jerry, P.-Y. Chen, J.~Zhang, P.~Sharma, K.~Ni, S.~Yu, and S.~Datta,
  ``Ferroelectric fet analog synapse for acceleration of deep neural network
  training,'' in \emph{2017 IEEE International Electron Devices Meeting
  (IEDM)}, 2017, pp. 6.2.1--6.2.4.

\bibitem{ch8fe3}
J.~Yoo, H.~Song, H.~Lee, S.~Lim, S.~Kim, K.~Heo, and H.~Bae, ``Recent research
  for hzo-based ferroelectric memory towards in-memory computing
  applications,'' \emph{Electronics}, vol.~12, no.~10, p. 2297, 2023.

\bibitem{ch8rram4}
J.~Woo, K.~Moon, J.~Song, S.~Lee, M.~Kwak, J.~Park, and H.~Hwang, ``Improved
  synaptic behavior under identical pulses using alox/hfo2 bilayer rram array
  for neuromorphic systems,'' \emph{IEEE Electron Device Letters}, vol.~37,
  no.~8, pp. 994--997, 2016.

\bibitem{ch8neurosim1}
P.-Y. Chen, X.~Peng, and S.~Yu, ``Neurosim+: An integrated device-to-algorithm
  framework for benchmarking synaptic devices and array architectures,'' in
  \emph{2017 IEEE International Electron Devices Meeting (IEDM)}.\hskip 1em
  plus 0.5em minus 0.4em\relax IEEE, 2017, pp. 6--1.

\bibitem{ch7ur1.ted}
D.~Lane, P.~Hodgson, R.~Potter, R.~Beanland, and M.~Hayne,
  ``\uppercase{ULTRARAM}: toward the development of a iii--v semiconductor,
  nonvolatile, random access memory,'' \emph{IEEE Transactions on Electron
  Devices}, vol.~68, no.~5, pp. 2271--2274, 2021.

\bibitem{mypaper_uredtm}
A.~Kumar, M.~Ehteshamuddin, A.~Bulusu, S.~Mehrotra, and A.~Dasgupta, ``A
  physics-based compact model for ultraram memory device,'' in \emph{2024 8th
  IEEE Electron Devices Technology and Manufacturing Conference (EDTM)}, 2024,
  pp. 1--3, doi: 10.1109/EDTM58488.2024.10512293.

\bibitem{mypaper_urdrc}
A.~Kumar and A.~Dasgupta, ``Compact modeling of compound semiconductor memory
  ultraram: A universal memory device,'' in \emph{2024 Device Research
  Conference (DRC)}, 2024, pp. 1--2, doi: 10.1109/DRC61706.2024.10605295.

\bibitem{ch8dnn4}
X.~Peng, S.~Huang, H.~Jiang, A.~Lu, and S.~Yu, ``Dnn+neurosim v2.0: An
  end-to-end benchmarking framework for compute-in-memory accelerators for
  on-chip training,'' \emph{IEEE Transactions on Computer-Aided Design of
  Integrated Circuits and Systems}, vol.~40, no.~11, pp. 2306--2319, 2021, doi:
  10.1109/TCAD.2020.3043731.

\bibitem{ch7ur2.aem}
P.~D. Hodgson, D.~Lane, P.~J. Carrington, E.~Delli, R.~Beanland, and M.~Hayne,
  ``\uppercase{ULTRARAM}: A low-energy, high-endurance, compound-semiconductor
  memory on silicon,'' \emph{Advanced Electronic Materials}, vol.~8, no.~4, p.
  2101103, 2022.

\bibitem{ch7tbrt3}
D.~Lane and M.~Hayne, ``Simulations of resonant tunnelling through inas/alsb
  heterostructures for \uppercase{ULTRARAM} memory,'' \emph{Journal of Physics
  D: Applied Physics}, vol.~54, no.~35, p. 355104, 2021.

\bibitem{ch7ur3.edtm}
D.~Lane, P.~Hodgson, R.~Potter, and M.~Hayne, ``Demonstration of a fast,
  low-voltage, \uppercase{III-V} semiconductor, non-volatile memory,'' in
  \emph{2021 5th IEEE Electron Devices Technology \& Manufacturing Conference
  (EDTM)}.\hskip 1em plus 0.5em minus 0.4em\relax IEEE, 2021, pp. 1--3.

\bibitem{vgg8_ref}
K.~Simonyan and A.~Zisserman, ``Very deep convolutional networks for
  large-scale image recognition,'' in \emph{International Conference on
  Learning Representations}, 2015.

\bibitem{vgg8_code}
M.~A. Rasslan, ``Alexnet, vgg16, and vgg8 on cifar-10,'' Kaggle Notebook, 2025,
  \url{https://www.kaggle.com/code/mennaalaarasslan/alexnet-vgg16-and-vgg8-on-cifar-10}.

\bibitem{myjap_UR}
A.~Kumar, M.~Dar, P.~Hodgson, D.~Lane, P.~Carrington, E.~Delli, R.~Beanland,
  S.~Mehrotra, M.~Hayne, and A.~Dasgupta, ``Physics, modeling, and benchmarking
  of ultraram: A compound semiconductor-based memory device,'' \emph{Journal of
  Applied Physics}, vol. 138, no.~9, 2025, doi: 10.1063/5.0269780.

\bibitem{ch8memristor1}
S.~H. Jo, T.~Chang, I.~Ebong, B.~B. Bhadviya, P.~Mazumder, and W.~Lu,
  ``Nanoscale memristor device as synapse in neuromorphic systems,'' \emph{Nano
  letters}, vol.~10, no.~4, pp. 1297--1301, 2010.

\bibitem{ch8epiram}
S.~Choi, S.~H. Tan, Z.~Li, Y.~Kim, C.~Choi, P.-Y. Chen, H.~Yeon, S.~Yu, and
  J.~Kim, ``Sige epitaxial memory for neuromorphic computing with reproducible
  high performance based on engineered dislocations,'' \emph{Nature materials},
  vol.~17, no.~4, pp. 335--340, 2018.

\bibitem{ch8_benchrram1}
S.~Park, A.~Sheri, J.~Kim, J.~Noh, J.~Jang, M.~Jeon, B.~Lee, B.~Lee, B.~Lee,
  and H.-J. Hwang, ``Neuromorphic speech systems using advanced reram-based
  synapse,'' in \emph{2013 IEEE International Electron Devices Meeting}.\hskip
  1em plus 0.5em minus 0.4em\relax IEEE, 2013, pp. 25--6.

\end{thebibliography}

\end{document}